\documentclass[doublecol]{epl2} 
\usepackage{amsmath}
\usepackage{graphicx}
\usepackage{psfrag}
\title{A Generalized Theory of DNA Looping and Cyclization}

\author{David P. Wilson\inst{1} \and Alexei V. Tkachenko\inst{2} \and Jens-Christian Meiners\inst{1,3}}
\shortauthor{D. P. Wilson \etal}

\institute{                    
\inst{1} Department of Physics, University of Michigan, 450 Church St., Ann Arbor, MI 48109 USA  \\
\inst{2} Center for Functional Nanomaterials, Brookhaven National Laboratory, Bldg 735, Upton NY 11973 USA\\
\inst{3} LSA Biophysics, University of Michigan, 930 N University Ave, Ann Arbor, MI 48109 USA 
}

\abstract{We have developed a generalized semi-analytic approach for efficiently computing cyclization and looping $J$ factors of DNA under arbitrary binding constraints.  Many biological systems involving DNA-protein interactions impose precise boundary conditions on DNA, which necessitates a treatment beyond the Shimada-Yamakawa model for ring cyclization.  Our model allows for DNA to be treated as a heteropolymer with sequence-dependent intrinsic curvature and stiffness.  In this framework, we independently compute enthlapic and entropic contributions to the $J$ factor and show that even at small length scales $(\sim  \ell_{p})$ entropic effects are significant.  We propose a simple analytic formula to describe our numerical results for a homogenous DNA in planar loops, which can be used to predict experimental cyclization and loop formation rates as a function of loop size and binding geometry. We also introduce an effective torsional persistence length that describes the coupling between twist and bending of DNA when looped.}

\begin{document}
\maketitle

\section{Introduction}
\begin{figure}
\includegraphics{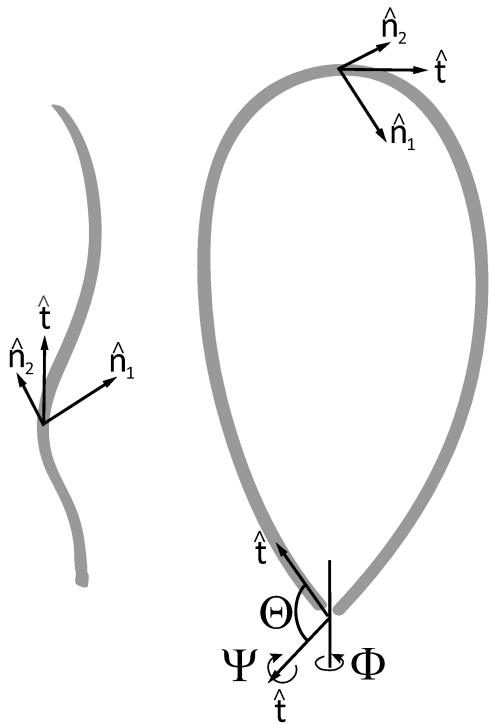}
\caption{\small Representation of the local basis vectors along the DNA in the open and looped states, respectively.  In the looped state we prescribe the end point locations through a set of spatial coordinates $(x,y,z)$ and angles $(\Theta, \Phi, \Psi)$ between end point tangent vectors.  Note the directions of $\hat{n}_{1}, \hat{n}_{2}$ are determined by the open state body fixed frame, rather than the looped state space-curve.  Also note that the circular Ring corresponds to $\Theta = \Phi = 0$.  Phasing of the two end points is represented by  $\Psi$, and is often due to a mismatch in the helical repeat of $10.5$ base pairs.}
\label{coords}
\end{figure}

Calculating the probability that contact will occur between two distant ends of a polymer under prescribed orientations is a long-standing question of considerable significance in polymer physics. This problem was rigorously defined in the context of polyelectrolyte condensation as the ratio of equilibrium constants for cyclization and bimolecular association by introduction of the Jacobson Stockmayer $(J)$ factor \cite{Jacobson1950}.  Yamakawa and Stockmayer  expanded on this work using the Kratky-Porod wormlike chain model (WLC) to compute the $J$ factor of angle-independent DNA ring-closure probabilities \cite{Yamakawa1972}.  Shimada and Yamakawa then included twist alignment of the end points \cite{Shimada1984b}, known as phasing, to explain the  measured oscillatory cyclization rates by Shore and Baldwin on DNA shorter than $500$ base pairs \cite{Shore1981}.  Shimada and Yamakawa calculated the $J$ factor for the Ring and and unconstrained loop, by treating DNA as a homo-polymer with coincident end points and parallel tangent vectors, as well as with coincident end points with unconstrained tangent vectors, respectively, see fig.~\ref{coords}.  Our work generalizes this closure probability to include arbitrary end point locations, binding orientations, sequence-dependent curvature and elasticity. \\
\indent We numerically calculate $J$ factors based on a semi-analytic formulation that includes specified end point locations and orientations of the DNA.  This formulation goes beyond the homogeneous straight elastic rod \cite{Balaeff1999,Balaeff2004,Balaeff2006} by including as inputs intrinsic curvature and stiffness based upon sequence-dependent effects.  While Monte Carlo methods have been successfully used to compute $J$ factors \cite{Towles2009,Pavone2001}, in general, it is difficult to separate out the individual effects of curvature and stiffness given they are very computationally taxing; by contrast, our computation of the $J$ factor based on a desired equilibrium shape takes only minutes on a MacBook Pro. 

Our model independently calculates enthalpic and entropic contributions to the free energy of the DNA loop.  The numerical results show that boundary-condition dominated entropic contributions are important even for very short DNA on the order a persistence length $(\ell_{p})$.  Within a cell, DNA is normally constrained by histones and other binding constraints, leaving this length scale as the typical size of locally fluctuating DNA. 

Many DNA-binding proteins impose very specific boundary conditions on DNA loop formation.  Previous results have shown that boundary condition constraints on the DNA end points play a significant role in the facilitation of loop formation \cite{Segall2006}.  Boundary conditions have also been suggested by Tkachenko \cite{Tkachenko2007} as an explanation for the striking disagreements between the cyclization rates measured by Du et al. \cite{Du2005,Du2005b} and Cloutier et al. \cite{Cloutier2004,Cloutier2005}.  Therefore, any useful model for these interactions must accommodate such arbitrary boundary conditions.  Thus, the $J$ factor framework gives quantitative insights into the mechanics of protein-mediated DNA loop formation and is important for multi-scale models of larger DNA-protein assemblies such as chromatin and
nucleosomes.

\section{Theory}
To model the enthalpic and entropic contributions to the looping $J$ factor, we use a coarse-grained elastic rod model for the DNA polymer to calculate the Hamiltonians describing thermal fluctuations about the open $(H^{o})$ and looped $(H^{\ell})$ states.  The $J$ factor is then calculated by comparing the probability densities of finding the DNA in a configuration that corresponds to the looped state with enforced boundary conditions for the end points, described by three angles $(\Theta, \Phi, \Psi)$ and three positions $(x,y,z)$ (see fig.~\ref{coords}), to that of the open state without such constraints
\begin{align}
J & = \frac{8 \pi^2 \int [d\xi_i] e^{-\beta H^{\ell}(\xi_{i})} \delta^3(\vec{u}(L))\delta(\theta_1(L)) \delta(\theta_2(L))\delta(\psi(L))}{\int [d\xi_i] e^{-\beta H^{o}(\xi_{i})}}.
\label{Jintegral}
\end{align}
The integration is over the amplitudes $d \xi_{i}$ of the normal modes with eigenvalue $\lambda_{i}$ of the respective Hamiltonians, and $\beta = 1/\left(k_{B}T \right)$, the inverse product of the Boltzmann constant $k_{B}$, and temperature $T$.  The DNA is parameterized by arc length parameter $s$, where $s=0$ and $s=L$ are taken to be the end points.  The endpoint tangent vectors have three angular constraints $\theta_{1}(L), \theta_{2}(L), \psi(L)$ and a relative displacement vector $\vec{u}$ which are imposed by $\delta^{3}(\vec{u})$, $\delta(\theta_{1,2})$ and $\delta(\psi)$, respectively.

The open state is characterized by three input local curvature components $\vec{\kappa}^{o} = (\kappa_{1}^{o}(s), \kappa_{2}^{o}(s), \tau^{o}(s))$, which represent intrinsic curvature caused by sequence-dependence.  We include as inputs two bending persistence lengths $\ell_{1}(s), \ell_{2}(s)$, corresponding to bending elasticity along the major and minor grooves of DNA, respectively, as well as a torsional persistence length $\ell_{\tau}(s)$.  

The three equilibrium looped state curvature components $\vec{\kappa}^{\ell} = (\kappa_{1}^{\ell}(s), \kappa_{2}^{\ell}(s), \tau^{\ell}(s))$ are found by minimizing the strain energy of DNA under specified orientations and positions of the endpoint tangent vectors, while tracking the DNA cross sections, as demonstrated by Goyal et al. \cite{Goyal2005}.  This tracking allows the use of the body fixed vectors $(\hat{t}(s),\hat{n}_{1}(s),\hat{n}_{2}(s))$ for a basis to define angular deformations  $(\theta_{1}(s),\theta_{2}(s),\psi(s))$ from the equilibrium open and looped states.  The angles $\theta_{1,2}(s)$ are defined as rotations about the two open state normal vectors and $\psi(s)$ is defined as a rotation about the tangent vector of the open state. 

The deformation induced curvatures $\tilde{\kappa}^{o,\ell} = (\tilde{\kappa}_{1}^{o,\ell}(s), \tilde{\kappa}_{2}^{o,\ell}(s), \tilde{\tau}^{o,\ell}(s))$ are computed separately for the open $(\tilde{\kappa}^{o})$ and looped $(\tilde{\kappa}^{\ell})$ states, respectively.  The looped state Hamiltonian is expressed as $H^{\ell} = E^{\ell} + \delta H (\theta_{1},\theta_{2},\psi)$, where $E^{\ell}$ is the strain or enthalpic energetic cost of loop formation.  The open state equilibrium is the intrinsic curvature induced by sequence-dependence.  As DNA is in an aqueous solution, its mobility is severely overdamped and the kinetic energy contributions to the Hamiltonian are neglected.  The deformation Hamiltonian is then
\begin{align}
\beta H^{o,\ell}  =& \frac{1}{2} \int_0^L d{s} \left(\tilde{\kappa}^{o,\ell} -\vec{\kappa}^o \right)^T B(s) \left(\tilde{\kappa}^{o,\ell} -\vec{\kappa}^o \right) 
\label{eq:Hamiltonian}
\end{align}
where $B(s)$ is the stiffness tensor, which we take to be diagonal with components $\ell_{1}(s),\ell_{2}(s)$ and $\ell_{\tau}(s)$.  The curvatures components $\tilde{\kappa}$ in eq.~\ref{eq:Hamiltonian} are organized into three groups based on their order of deformation variables $(\theta_{1}, \theta_{2}, \psi)$.  The zeroth order terms represent strain energy of loop formation.  The first order terms define the equilibrium conditions, and will vanish.  The second order terms determine the normal modes, while the higher order terms are neglected.  The equilibrium planar deformation curvatures that we will work with in this paper are
\begin{align}
(\tilde{\kappa}_{1}^{2} + \tilde{\kappa}_{2}^{2}) =& \left( \kappa_{2}^{2} + 2 \kappa_{2} \theta_{2}' \right) + \left(  (\theta_{1}'^{2} - \kappa_{2}^{2}\theta_{1}^{2}) + \theta_{2}'^{2} \right) \\
\tilde{\tau}^{2} =&  \left( \psi' - \kappa_{2} \theta_{1}\right)^{2}.
\end{align}
Here we have assumed the DNA to be isotropic in bending stiffness, $\ell_{p} = \ell_{1} = \ell_{2}$, and intrinsically straight.  The curvature components contained in the Hamiltonian are made non-dimensional by scaling with the overall DNA length, $L$. 

When constructing the Hamiltonian, the Galerkin method is used to numerically solve for the normal modes of the open and looped states.  Each deformation variable $(\theta_{1}(s), \theta_{2}(s), \psi(s))$ is expanded in terms of $N$ orthogonal comparison functions, which are then used to create a $3N \times 3N$ Hamiltonian matrix for the open $\mathcal{H}^{o}$ and the looped $\mathcal{H}^{\ell}$ states.  The comparison functions satisfy the angular boundary constraints imposed by $\delta (\theta_{1}(L)),\delta (\theta_{2}(L)), \delta (\psi(L))$ in eq.~\ref{Jintegral}.  The remaining looped boundary condition $\delta^{3}(\vec{u})$ are satisfied by Fourier expanding the delta functions, and then integrating over the eigenvector amplitudes $\xi_{i}$, leading to the constraint matrix $V$.  An additional integration for the open state is required to cover the modes  which cause displacements of the end points.  The $J$ factor can then be expressed as 
\begin{align}
J =& \frac{1}{ \ell_p^3} \sqrt{ \frac{\det \mathcal{H}^o}{ 2 \pi^{3} \det \mathcal{H}^\ell  \det V}\left( \frac{\ell_p}{L}\right)^{11}}  e^{-\frac{1}{2}\frac{\ell_p}{L} \int \kappa_p^2 \text{ds} - \frac{L}{4\ell_p} } \nonumber \\ 
	=& \Lambda (\Theta) \exp \left(-\frac{1}{2}\frac{\ell_p}{L} \int \kappa_p^2 \text{ds} -  \frac{L}{4 \ell_p} \right), \label{eq:Jfunction}
\end{align}
which is a product of two functions, one describing the entropic contributions, henceforth referred to as the entropic coefficient $\Lambda(\Theta)$, and the exponential term containing the enthalpic contributions.  The lowest eigenmodes of the Hamiltonians converge in the limit of large $N$.  The ratio of Hamiltonian determinants is finite because higher spatial frequency modes are less sensitive to the curvature of the shape.  After the $M^{th}$ eigenmode, the ratio of eigenvalues converges
\begin{align}
\frac{\det \mathcal{H}^o}{\det \mathcal{H}^{\ell}} = \frac{\lambda_{1}^{o} \cdots \lambda_{M}^{o}}{\lambda_{1}^{\ell} \cdots \lambda_{M}^{\ell}}.
 \end{align}
The ratio of eigenvalues describes how the space accessible to thermal fluctuations of the DNA is reduced upon loop formation, which in turn allows us to quantify the entropic change of the system.

\section{Results}
The results presented here are for near planar DNA loops with coincident end points and arbitrary loop tangent angle $\Theta$.  In this paper we present results for DNA of the length $50 nm$ or approximately $14$ helical repeats, so we will assume $\ell_{1} = \ell_{2}$.  We treat the DNA as a homogeneous polymer with bending and torsional persistence lengths of $50$ nm and $75$ nm, respectively.\cite{Strick1996,Hagerman1988,Baumann1997}.  While DNA is a heteropolymer with anisotropic bending persistence lengths $\ell_{1}(s) \neq \ell_{2}(s)$, this anisotropy largely averages out after a few helical repeats $(~10.5 \text{ base pairs})$ as demonstrated by Kehrbaum and Maddocks \cite{Kehrbaum2000}.  

\begin{figure}
\psfrag{0}[][c]{$0$}
\psfrag{0.7854}[][c]{$\frac{1}{4}\pi$}
\psfrag{1.5708}[][c]{$\frac{1}{2}\pi$}
\psfrag{2.3562}[][c]{$\frac{3}{4}\pi$}
\psfrag{3.1416}[][c]{$\pi$}
\psfrag{xlabel}{$\Theta(\textrm{radians})$}
\psfrag{ylabel}{Molarity}
\psfrag{title}[c][c]{DNA Length $= 50$ nm}
\psfrag{first}{$\gamma(\Theta)$}
\psfrag{second}{$\Lambda(\Theta)$}
\onefigure{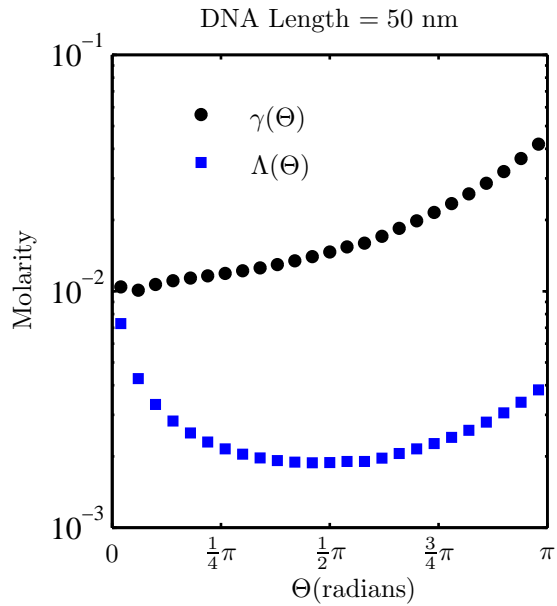}

\caption{ \small The entropic coefficient $\Lambda(\Theta)$ is largely dominated by contributions from the lowest eigenmode of the loop, $\lambda_{1}^{\ell}(\Theta)$.  To illustrate this dependence, we write $\Lambda(\Theta) = f(\lambda_{1}^{\ell}(\Theta)) \gamma(\Theta)$, where $f(\lambda_{1}^{\ell})$ contains only the contributions of the lowest eigenmode $\lambda_{1}^{\ell}$.  The function $f(\lambda_{1}^{\ell})$ is represented by the bracketed quantity in eq.~\ref{eq:Bessel}, and $\gamma(\Theta)$ is given in eq.~\ref{eq:Gamma}.  It is then clear that $\gamma(\Theta)$ is a slowly varying function on the interval $(\Theta = 0)$ to $(\Theta = 0.54 \pi)$, and then steadily increases on the interval from $\Theta = 0.54 \pi$ to Hairpin $(\Theta = \pi)$.  The shift in behavior of $\gamma(\Theta)$ occurs after the lowest eigenmode changes from symmetric to antisymmetric.  Even for relatively short DNA $\Lambda(\Theta)$ is shown to effect the $J$ factor by an order of magnitude in fig.~\ref{plotJ}.  Note the dimensions are in molarity rather than concentration as in eq.~\ref{eq:Jfunction}.} 
\label{plotEC}
\end{figure}

The entropic coefficient $\Lambda(\Theta)$ is computed for a torsionally unconstrained DNA loop with overall length $L = \ell_{p}$ and loop tangent angles ranging from a Ring $\Theta = 0$, to a Teardrop  $\Theta \sim 0.54 \pi $, to a Hairpin $\Theta = \pi$ and is given in fig.~\ref{plotEC}.  The lowest eigenvalue of the in-plane loops can be well approximated as $\lambda^{\ell}_{1} = 2 \pi \Theta$.  Factoring this contribution from $\Lambda(\Theta)$ reveals a slowly varying function  $\gamma(\Theta)$.  We are able to fit our numerical results to within $1\%$ by using a modified Bessel Function
\begin{align}
J(\Theta)   = &  \left[ I_0(2 \pi \Theta) e^{-2 \pi \Theta}\right]  \gamma(\Theta) \frac{1}{\ell_p^3} \left(\frac{\ell_p}{L} \right)^{11/2},  \nonumber \\
& \quad \times \exp\left(-\frac{\ell_p}{L} E(\Theta) - \frac{L}{4\ell_p} \right)  \label{eq:Bessel} \\
\gamma(\Theta)  = &  365 \Theta^{2} - 525 \Theta + 32 \pi^{3}, \label{eq:Gamma} \\
E(\Theta) = &\frac{1}{2}\int \kappa_p^2 \text{ds}  = 2.02 (\Theta - 0.54 \pi)^{2} + 14.05  \label{eq:Etheta} , 
\end{align}
where $\Theta$ is the loop tangent angle in radians.  The fit is accurate for all angles $\Theta$ although the dimensional scaling of eq.~\ref{eq:Bessel} needs to be modified to $\left(\ell_p/L \right)^{6}$ when $\Theta = 0$, as the ring has a zero mode \cite{Shimada1984b}.   The unconstrained loop by contrast has dimensional scaling of $\left(\ell_p/L \right)^{5}$, due to integrating over the orientations of the tangent vectors.

\begin{figure}
\psfrag{0}[][c]{$0$}
\psfrag{0.7854}[][c]{$\frac{1}{4}\pi$}
\psfrag{1.5708}[][c]{$\frac{1}{2}\pi$}
\psfrag{2.3562}[][c]{$\frac{3}{4}\pi$}
\psfrag{3.1416}[][c]{$\pi$}
\psfrag{xlabel}{$\Theta(\textrm{radians})$}
\psfrag{ylabel}{$J(M)$}
\psfrag{title}[c][c]{DNA Length $= 50$ nm}
\psfrag{first}{$\Lambda(\Theta=0)$}
\psfrag{second}{$\Lambda(\Theta)$}
\psfrag{third}{$\langle\Lambda(\Theta)\rangle$}
\onefigure{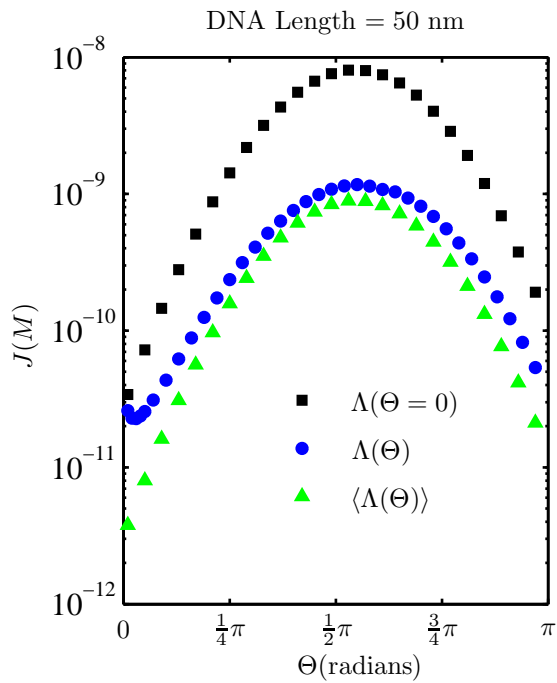}
\caption{ - \small A comparison of entropic coefficient effects on the $J$ factors computed using only enthalpic considerations versus a full treatment of enthalpic and entropic considerations.  In this way we demonstrate via several orders of magnitude difference that the entropic changes are vital to the calculation of the $J$ factor.  Small angles and Hairpin structures are poorly described by the enthalpic only extrapolations of the $J$ factor.  These results are for DNA of length $50$ nm and increase in difference as length is increased.  }
\label{plotJ}
\end{figure}

The Teardrop  shape has the lowest strain (enthalpic) energy, $14.4 \frac{\ell_{p}}{L} \ k_{B} T$, of any of the in-plane shapes, and is where the endpoint curvatures of the loop vanish. Enthalpic considerations demonstrate which loop tangent angle $\Theta$ will produce the maximum $J$ factor, although for small angles, as well  Hairpin structures, entropic considerations are required to demonstrate the absolute behavior of the  $J$ factor, as seen in fig.~\ref{plotJ}.

Shimada and Yamakawa provided two special cases for the in-plane $J$ factors \cite{Shimada1984b}: the Ring defined by aligned tangents, $\Theta = 0$, and the unconstrained loop.  We have reproduced the Ring and unconstrained results to within $0.01\%$.  The unconstrained loop formation assumes that $E(\Theta)$ given above in eq.~\ref{eq:Etheta} is symmetric about the Teardrop.  As most biologically relevant cases do not fit neatly into one of these special cases, our generalized results allow a more accurate prediction of the $J$ factor.  To illustrate the effect of angular dependence on the $J$ factor, we plot three $J$ factors with different entropic coefficients $\Lambda$ in fig.~\ref{plotJ}.  Two of these $J$ factors have constant entropic coefficients $\Lambda \neq \Lambda(\Theta)$, that of the Ring and unconstrained loop, while allowing the normal angular dependence of the enthalpic contributions $E(\Theta)$.  Thus fig.~\ref{plotJ} demonstrates that the changes to the entropic contributions as a function of $\Theta$ are critical in the $J$ factor calculation.
\begin{figure}
\psfrag{0}[][c]{$0$}
\psfrag{0.7854}[][c]{$\frac{1}{4}\pi$}
\psfrag{1.5708}[][c]{$\frac{1}{2}\pi$}
\psfrag{2.3562}[][c]{$\frac{3}{4}\pi$}
\psfrag{3.1416}[][c]{$\pi$}
\psfrag{xlabel}{$\Theta(\textrm{radians})$}
\psfrag{ylabeltop}{$\ell_{\tau}^{*}(\ell_{p})$}
\psfrag{ylabelbottom}[][c]{$\alpha$}
\psfrag{(a)}{$\textbf{(a)}$}
\psfrag{(b)}{$\textbf{(b)}$}
\onefigure{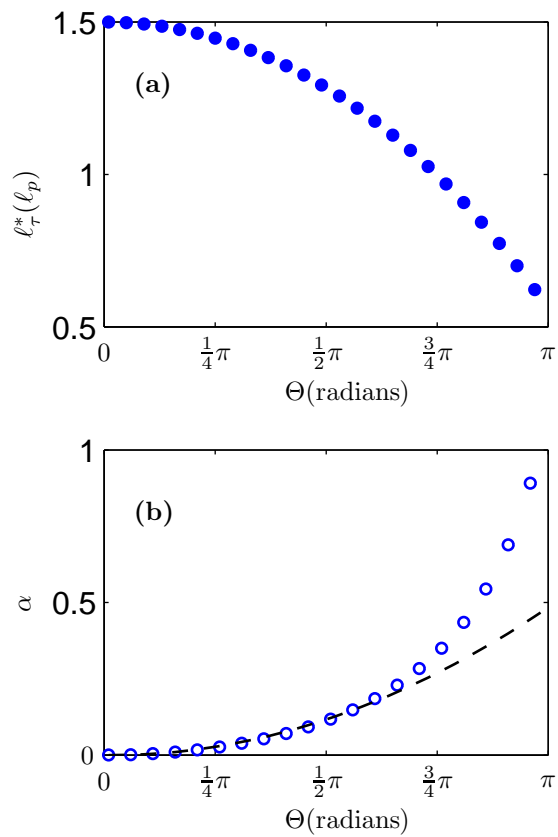}
\caption{\small \textbf{(a)} The effective torsional persistence length, $\ell_{\tau}^{\mathrm{*}}$ in units of $\ell_{p}$ as a function of loop formation angle, $\Theta$.  The Ring has pure torsional modes with stiffness $\lambda_{i} \frac{\ell_{\tau}}{L}$  and as $\Theta$ increase the bending and torsional modes become coupled, reducing the effective torsional persistence length. \textbf{(b)} The torsion-bending coupling $\alpha(\Theta)$ shown as open circles, is quadratic from the Ring to the Teardrop as seen by the dashed line.  From the Teardrop to the Hairpin, $\alpha(\Theta)$ is cubic in $\Theta$.}
\label{plot4}
\end{figure}

The general DNA-protein complex has spatially separated end points as well as prescribed angles $(\Theta, \Phi, \Psi)$ which can be obtained from  DNA-protein co-crystals with LacI protein serving as the canonical example \cite{Lewis1996}.  The extrapolation from the orientation averaged loop is better than the ring, except for small angles $\Theta$.  For all angles, the Bessel function given in eq.~\ref{eq:Bessel} is an excellent fit, with a maximum error of less than $1\%$ for all $\Theta$.

Computing the torsionally constrained $J$ factor by including $\delta(\psi)$ in eq.~\ref{Jintegral} allows a determination of an effective torsional persistence length $\ell_{\tau}^{*}$.  The coupling of torsion and bending elasticity can be computed as
\begin{align}
\ell_{\tau}^{*} &= 2 \pi \left(\frac{J(\Theta)_{\psi = 0}}{J(\Theta)_{\psi \neq 0}} \right)^{2}.
\end{align}
where $J(\Theta)_{\psi = 0}$ and $J(\Theta)_{\psi \neq 0}$ are the torsionally constrained and unconstrained $J$ factors, respectively.  The effective torsional persistence length represents the conversion between twist and writhe.  

The effective torsional persistence length can be written as a torsional and bending spring in series 
\begin{align}
\frac{1}{\ell_{\tau}^{*}} & = \frac{1}{\ell_{\tau}} + \alpha(\Theta) \frac{1}{\ell_{p}},
\end{align}
where all of the angular dependence is given by $\alpha$.  In fig.~\ref{plot4}, it is clear that $\alpha(\Theta)$ has as simple quadratic dependence up until the Teardrop shape and afterwards becomes cubic in $\Theta$
\begin{align}
\alpha(\Theta)  &= \frac{1}{2 \pi^{2}} \Theta^{2} - \frac{1}{6 \pi^{3}} \Theta, \quad 0 \leq \Theta \leq 0.55 \pi,  \\
\alpha(\Theta)  &=0.42 \Theta^{3} - 2.55 \Theta^{2} + 5.46 \Theta - 3.87,  0.55 \pi \leq \Theta  \leq \pi \nonumber \\
\end{align}

\section{Conclusion}
We have developed a generalized approach for computing $J$ factors of arbitrary loop shapes, which may include sequence-dependent stiffness and curvature.  We have shown that the $J$ factor varies strongly for near planar loop shapes as a function of loop tangent angle $\Theta$ for intrinsically straight DNA with isotropic bending stiffness.  The in-plane $J$ factors can be well fit with analytic functions for all $\Theta$.  We have defined an effective torsional persistence length $\ell_{\tau}^{*}$ and subsequence torsion-bending coupling $\alpha(\Theta)$ which are shown to vary significantly as a function of loop formation angle, $\Theta$.  Finally, our calculation is computationally very quick, taking only a few minutes per $J$ factor for any set of input boundary conditions.  
\acknowledgments
The authors would like to thank Noel Perkins, Sachin Goyal, Todd Lillian, Gerhard Blab, Krishnan Raghunathan, Patrick Koehn, Yih-Fan Chen and Margaret Wilson for their many helpful conversations.  This work was partially funded through grant GM 65934 from the National Institutes of Health.

\bibliographystyle{unsrt}
\bibliography{BLR}
\end{document}